\documentstyle[buckow,latexsym,amssymb]{article}

\def\beq{\begin{equation}}
\def\eeq{\end{equation}}
\def\ii{{\rm i}}
\def\ee{\mathrm{e}}

\newcommand{\eq}[1]{Eq.~(\ref{#1})}

\begin {document}
\begin{flushright}
DFTT 38/2001 \\
 LAPTH-882/01 
 \end{flushright}
\def\titleline{
Fractional branes on ALE orbifolds
}
\def\authors{
M. Bill\'o\1ad, L. Gallot\2ad  and A. Liccardo \3ad }
\def\addresses{
\1ad
Dipartimento di Fisica Teorica - Univerisit\'a di Torino
and I.N.F.N. - Sezione di Torino\\
Via P. Giuria 1, 10125 Torino (Italy)\\
{\tt billo@to.infn.it} \\
\2ad Laboratoire d'Annecy-Le-Vieux de Physique Th\'eorique\\
LAPTH, CNRS, UMR 5108, Universit\'e de Savoie\\
Chemin de Bellevue, B.P. 110,
F-74941 Annecy-le-Vieux Cedex (France)
\\
{\tt gallot@lapp.in2p3.fr}\\
\3ad I.N.F.N. - Sezione di Napoli \\
Complesso Universitario di Monte Sant'Angelo,
Via Cintia ed. G,
80126 Napoli (Italy)\\
{\tt  Antonella.Liccardo@na.infn.it}
}
\def\abstracttext{
We derive the classical type IIB supergravity solution describing fractional D3-branes
transverse to a ${\mathbb C}^2/\Gamma$ orbifold singularity, for $\Gamma$ any
Kleinian ADE subgroup. This solution fully describes the ${\mathcal{N}}=2$
gauge theory with appropriate gauge groups and matter living on the branes, up
to non-perturbative instanton contributions.}
\large
\makefront

\section{Introduction}
One of the present major challenges in string theory is that of extending the
AdS/CFT duality to more general gravity/gauge theory relations, valid in
non-conformal situations. A  possibility is that of considering fractional
branes in singular spaces such as conifolds
(corresponding to ${\mathcal{N}}=1$ gauge theory) \cite{KN} or 
orbifolds (corresponding  to ${\mathcal{N}}=2$ 
gauge theory).
\par
Supergravity solutions dual to non-conformal theories are generically singular.
Some stringy mechanism 
(like the conifold deformation \cite{KS} or the enhan\c con \cite{JPP} 
mechanism in the orbifold cases) is expected to repair the singularities so as
to match the true IR behaviour of the field theory. 
\par
In this communication, based on \cite{noi}, we deal with ${\mathcal{N}}=2$ 
gauge theories arising from fractional brane configurations, 
where one  has good control on the
field theory side (through the Seiberg-Witten \cite{SW} exact low energy 
effective action) and on the string theory side (strings on an orbifold). This
allows for detailed comparisons at the perturbative level and (hopefully) at the
level of instanton corrections.
\par Fractional branes on the ${\mathbb C}^2/{\mathbb Z}_2$ orbifold have been
thoroughly investigated in the literature \cite{BVFLMP,PRZ}. By considering the more
general case of fractional brane configurations on ${\mathbb C}^2/\Gamma$, 
we find a very elegant agreement of the gravitational solutions, regulated by
appropriate enhan\c cons, with the perturbative treatment of the 
corresponding  gauge theories. We also get more insight on the role of the  
flux of the RR five-form.

\section{Closed and open strings on ${\mathbb C}^2/\Gamma$}
We consider type IIB string theory on a space-time of the form
${\mathbb R}^{1,5}\times {\mathbb C}^2/\Gamma$,
in which fractional D3-branes, transverse to the orbifold, are located at the
orbifold singularity $z^1=z^2=0$ (having indicated as $z^{1,2} \equiv x^{6,8}
+\ii x^{7,9}$ the coordinates of ${\mathbb C}^2$). $\Gamma$ is a
discrete subgroup of ${\rm SU}(2)$, acting on ${\mathbb C}^2$ by
\beq
\label{due}
g\in\Gamma: ~{z^1 \choose z^2} \mapsto {\mathcal Q}(g){z^1 \choose z^2}~,
\eeq 
with ${\mathcal Q}(g)\in \mathrm{SU}(2)$ the defining 2-dim representation; the
origin is clearly a fixed point.
The Kleinian subgroups $\Gamma$ are ADE-classified (for instance, the 
${\mathbb Z}_n$ cyclic groups correspond to $A_{n-1}$ algebras)
through the McKay
correspondence \cite{McKay}: in the Clebsh-Gordan series
\beq
\label{tre}
{\mathcal Q}\otimes {\mathcal D}_I = \oplus_J
\widehat A_{IJ}\, {\mathcal D}_J~,
\eeq
$\widehat A_{IJ}$  turns out to be the adjacency matrix of an
\emph{extended} Dynkin diagram of a simply-laced Lie algebra ${\mathcal
G}_\Gamma$. So, 
decomposing $I= (0,i)$, the irreducible representations ${\mathcal D}_i$
correspond to the nodes of the diagram, i.e., to the simple roots $\alpha_i$;
the trivial irrep ${\mathcal D}_0$ corresponds to the extra node in the extended
diagram, i.e., to $\alpha_0 \equiv - \sum_i d_i \alpha_i$. The dimensions $d_i$
of the irrespses correspond to the Dynkin labels (while of course $d_0=1$).
It is important in the following the fact that the extended Cartan matrix
$\widehat C_{IJ} \equiv \alpha_I\cdot \alpha_J = 2\, \delta_{IJ} - \widehat
A_{IJ}$ is semidefinite positive, and admits the null eigenvector $d_I$:
$\widehat C_{IJ} d_J=0$.
\par
The orbifold ${\mathbb C}^2/\Gamma$ is singular at the origin. Resolving this
singularity in a canonical way produces \cite{Kronheimer} an Asymptotically 
Locally Euclidean (ALE) space ${\mathcal M}_\Gamma$. 
This space has a non-trivial middle homology $H_2({\mathcal M}_\Gamma)$,  
generated by two-spheres $e_i$ arising in the resolution. The intersection
form of these ``exceptional'' cycles is the (non-extended) Cartan matrix of 
${\mathcal G}_\Gamma$: $e_i\cdot e_j = -C_{ij}$. By Poincar\'e duality, the ALE
space possesses a set of anti-self-dual two-forms $\omega^i$ such that
\beq
\label{quattro}
\int_{e_i}{\omega^j} =
\delta^{{i}}_{{j}}
\hskip 0.1cm \leftrightarrow
\int_{{\rm ALE}} {\omega^j}\wedge
{\omega^j} = - {(C^{-1})^{ij}}~.
\eeq
\par
Closed type IIB strings in this orbifold background 
admit an exact SCFT description \cite{Aleman}; in the untwisted sectors
only $\Gamma$-invariant states are retained, and there are additional states
from the twisted sectors. The twisted sectors are in
correspondence with the conjugacy classes of $\Gamma$; in these sectors the
momentum has no components along the directions in which $\Gamma$ acts, so the
twisted fields have a 6-dimensional dynamics and are stuck at the fixed point
of the orbifold. Viewing ``geometrically'' the orbifold
background as the singular limit of the ALE space for vanishing two-cycles,
the low-energy supergravity theory can be obtained \`a la
Kaluza-Klein: in particular, decomposing the closed string massless fields on 
the two-forms $\omega^i$ dual to the cycles reproduces the spectrum of twisted
fields. In our solution, the following twisted scalar fields%
\footnote{In the closed string theory, the $b_i$ fields should be periodic of period $1$,
as world-sheet instantons on the vanishing cycle $e_i$ contribute 
$\exp (2\pi\ii
b_i)$ to the partition function. At the supergravity level, this symmetry is not
explicit.
}will be crucial:
\beq
\label{cinque}
b_i \equiv {1\over 2\pi} \int_{e_i} B_2~,\hskip 0.5cm
c_i \equiv {1\over 2\pi} \int_{e_i} C_2~.
\eeq
\par
Reducing type IIB sugra on the singular orbifold, 
one arrives to the bulk supergravity action (restricted to
the fields that will be turned on in the fractional brane solutions)       
\begin{eqnarray}
\label{sei}
S_{\rm b} & = & \frac{1}{2 \kappa^2} \Bigg\{\int d^{10} x~
\sqrt{-\det G}~ R
- \frac{1}{2} \int \Big[
\frac{1}{4\pi^2}
{d\gamma_i\wedge {}^* d\bar\gamma_j
\wedge\omega^i\wedge{}^*\omega^j}
\nonumber\\
& - &  \frac{\ii}{8\pi^2}C_4\wedge
{d\gamma_i
\wedge d\bar\gamma_j\wedge\omega^i\wedge\omega^j}
\,+\,\frac{1}{2}\, {\widetilde{F}}_{5}
\wedge {}^* {\widetilde{F}}_{5}\Big] \Bigg\}~,
\end{eqnarray}
where $\widetilde F_{5} \equiv dC_4+C_2\wedge dB_2$ is self-dual, and
{$\gamma_i \equiv c_i -\ii b_i$}. Also, one has 
$\kappa = 8 \pi^{7/2}g_s (\alpha')^2 = 2 \pi^{3/2} g_s$ (we use
$2\pi\alpha'=1$).
\par
When we introduce also D3 branes transverse to the orbifold space
\cite{quiver} 
the resulting open strings carry Chan-Paton factors that 
specify to which D3-branes they are attached. 
The group $\Gamma$ may act on the Chan-Paton factors as
well as on the string fields along the ${\mathbb C}^2$ directions.  
If we want a D3 to be located at a generic point $p$ inside the orbifold, then
the image branes at $gp$, $g\in \Gamma$, fill a complete orbit. The
Chan-Paton indices of open strings attached to such branes transform in the
regular representation of $\Gamma$; such configurations are called regular
or bulk branes. However, the regular representation decomposes as
${\mathcal R} = \oplus_I d_I\, {\mathcal D}_I$,
so a regular brane  at the origin of the orbifold (which is fixed
under $\Gamma$) it can decompose into
fractional branes. A fractional brane of type $I$ corresponds to Chan-Paton
indices in the irrep ${\mathcal D}_I$, and is stuck at the fixed point
\cite{Douglas:1996xg}.
In relation with the gauge theory living on the branes, the position of a
regular D3-brane inside ${\mathbb C}^2/\Gamma$ corresponds to the Higgs branch,
while the positions of its fractional constituents in the $x^4,x^5$ plane
correspond, as we will see, to the Coulomb branch.  
\par 
The fractional branes can be described as boundary states in the orbifold SCFT
\cite{DG,BCR}, but admit also a (quasi)-geometrical interpretation as D5 branes
wrapped on the vanishing cycles%
\footnote{The brane of type {$0$} is a D5 wrapped on
{$e_0=-\sum_i d_i e_i$} with a background gauge flux
$\int_{{e_0}} \mathcal{F} = 2 \pi$.} $e_i$.
By either methods, it is possible to infer the world-volume action for a
collection of $\{ m^I\}$ fractional branes of the various types:
\begin{equation}
\label{b2}
S_{\rm wv} = 
-\frac{T_3}{\kappa}\left\{ \int_{D3} \sqrt{-\mathrm{det} G }\,
{m^J b_J} + \int_{D3} C_{(4)}\, {m^J b_J} +
\int_{D3} {m^J {\mathcal A}_{\,4,J}}\right\}~.
\end{equation}
Here the tension has the usual expression $T_p = \sqrt{\pi}
(2\pi\sqrt{\alpha'})^{3-p}$, and the twisted RR 4-forms ${\mathcal A}_{4,J}$,
arising from $C_6 = {\mathcal A}_{4,J}\wedge\omega^J$, are dual in 6 dimensions
to the scalars $c_I$. Notice from \eq{b2} that the tension and charge of a brane
of type $I$ are proportional to the twisted field $b_I$. 

\section{The gauge theory on the brane}
The gauge theory living on a configuration of fractional D3 branes is determined
by the spectrum (and interactions) of the massless open string fields stretching
between such branes and surviving the orbifold projection, taking into
account also the action of $\Gamma$ on the Chan-Paton factors. The result is that
there is a ${\rm U}(m^I)$ ${\mathcal N}=2$ gauge multiplet from the strings stretching
between the $m^I$ branes of type $I$. Moreover, there are $\widehat A_{IJ}$ 
hypermultiplets in the $(m^I,\bar m^J)$ representation of ${\rm U}(m^I)$ and 
${\rm U}(m^J)$ respectively, from strings stretched between branes of type $I$ 
and $J$; namely, the number of hypermultiplets from such strings is the number of 
links between the nodes $I$ and $J$ in the extended
Dynkin diagram associated to $\Gamma$.     
The $\beta$-functions of the various ${\rm U}(m^I)$ gauge factors have only
one-loop perturbative contributions. 
The coefficients are given by
\begin{equation}
\label{b3}
2 m^I - \widehat A_{IJ}\, m^J = \widehat C_{IJ}\, m^J~,
\end{equation}
the positive contributions being from the gauge multiplet, the negative ones
from the matter hypermultiplets.
Thus, if we define a bare coupling $\bar g_I$ at an UV scale $\rho_0$, it
dimensionally transmutes into a RG invariant scale
$\Lambda_I$ and
the (complexified) coupling constants run as described by 
\begin{equation}
\label{b5}
{\tau_I}({\mu}) \equiv
\frac{4\pi\ii}{{g_{I}^2}} +\frac{{\theta_{I}}}{2\pi}
= \frac{\ii}{2\pi} {\widehat C_{IJ} m^J}\, \ln\frac{{\mu}}{{\Lambda_{I}}}~,
\hskip 0.8cm
\Lambda_I = \rho_0\ee^{-{8\pi^2\over \bar g_I^2 \widehat C_{IJ} m^J}}~.
\end{equation}
This ${\mathcal N}=2$ theory has a moduli space, parametrized by the v.e.v.'s of
the adjoint complex scalars in the vector multiplets, 
${\langle \phi\rangle} = \mathrm{diag}(a_1,\ldots,{a_{m_I}})$,
and by the masses $M_k$ ($k=1,\ldots N_{\mathrm{h}} =\widehat A_{IJ} m^J$) of
the hypers. A Seiberg-Witten description of the low energy effective action is
possible. The effective couplings have a perturbative part, which is determined
by one-loop diagrams with massive vector- and hyper-multiplets running in the
loop, plus non-perturbative corrections due to instantonic effects.
The analysis of \cite{PRZ} in the ${\mathbb C}^2/{\mathbb Z}_2$ case indicates
that in the limit of many branes the instantonic effects arise very suddenly
near the scales $\Lambda_I$ and, below $\Lambda_I$, the couplings $\tau_I$ tend
to stop running.

\section{The supergravity solution vs the gauge theory}
The equations of motion 
following from the bulk and boundary actions \eq{sei}
and \eq{b2} for the metric and the RR five-form
(for the explicit expressions  see \cite{noi})
 can be solved with the usual D3 brane ansatz in terms of a function
$H$ which depends upon all the transverse
coordinates, $z$ and $x^6,\ldots x^9$.
The other untwisted fields, in particular the dilaton $\varphi$ and the 
RR scalar $C_0$ are trivial. 
The only excited twisted fields are the
scalars $\gamma_i = c_i - \ii b_i$; they depend only on $z$. One finds that the
choice of constant dilaton and axion is consistent provided $\gamma_i(z)$ are
analytic functions. This being the case, one can also prove that half susies are
preserved in the bulk (see \cite{GMMP} in the case of a non-collapsing ALE).
\par 
Solving the eq. of motion for the
twisted scalars we find in fact the harmonic functions:
\begin{equation}
\label{solgamma}
{\gamma_I}({z}) = -\ii\, {g_s\over 2\pi}\, {\widehat C_{IJ}m^J}\,
\ln {\frac{z}{\Lambda_I}}~,
\end{equation}
where $\Lambda_I$ are introduced as IR (small $z$) regulators. We see that
these fields have exactly the same logarithmic running as the gauge couplings,
suggesting the correspondence 
\begin{equation}
\label{corrgamma}
\tau_I(z) \leftrightarrow - \gamma_I(z)/g_s~.
\end{equation}
In this perspective, $z$ corresponds to the complexified energy scale,
$\Lambda_I$ to the dynamically generated scales; moreover, the distance 
$|z|=\rho_0$ where the twisted fields attain their perturbative orbifold value
$b_I = d_I/|\Gamma|$ corresponds to the UV scale where the bare theory is
defined, with bare coupling $\bar g_I^2 = 4\pi g_s\,
|\Gamma|/d_I$.
\par
This RG/twisted field duality can be justified in the brane-probe approach.
Indeed, a single test brane of type $I$ placed at a point $z$ on the fixed 
plane in the background 
generated by a configuration of $\{m^I\}$ branes is BPS. The effective
two-derivative action for the U$(1)$ living on the test brane 
is found to be an ${\mathcal N}=2$ effective theory with coupling 
$\tau_I(z)$, $z$ being the position of the probe, where $\tau_I(z)$,
see Eqs. (\ref{solgamma},\ref{corrgamma}) is the twisted
scalar generated by the background. This is the same answer given by the
perturbative part of the Seiberg-Witten prepotential for this particular 
breaking of U$(m_I+1)$ to U$(1)\times {\rm U}(m_I)$.   
These  perturbative contributions arise from 1-loop diagrams in
which the massive vector- and hyper-multiplet fields run in the loop and 
correspond to loops of open strings stretched between the
test brane and  the background branes. Upon open/closed duality, 
these diagrams correpond to the tree-level 
propagation of massless closed strings, the twisted scalars, 
from the background branes to the probe, that is, to the evaluation of the probe
action in the classical background.
The situation is similar to the relations between RG flows and NS-NS tadpoles
found in different models \cite{BM,AA}.
\par
Besides the twisted fields, the fractional brane solution displays a non-trivial
metric and RR 4-form, parametrized by the function $H(z,x^i)$. 
The  effective tension and RR
charge of a type $I$  fractional branes, proportional to $b_I(z)$, attains its
orbifold value $d_I/|\Gamma|$ at the UV scale $\rho_0$, while it vanishes at the
enhan\c con radius $\Lambda_I$. The construction of the boundary action is 
justified at the orbifold point, but we assume we can trust it down to
$\Lambda_I$, where instanton effects become very suddenly important. So we
consider the fractional branes of each type%
\footnote{We consider configurations such that all but one twisted field run to
UV asymptotic freedom; then the remaining one, say $b_0$, which is not
independent, has the opposite running.} 
$i$ to be located on an enhan\c con shell of radius $\Lambda_i$ in the $z$-plane
(and those of type 0 at a point where $b_0=1$).
The solution of the RR 4-form and Einstein equation 
involves IR divergent
integrations, which are regulated via the 
enhan\c cons (recall that by
Seiberg-Witten analysis \cite{PRZ} the twisted fields $\gamma_i$ are expected to
basically stop running below their enhan\c con); see also \cite{Merlatti}. 
Above the highest enhan\c con, 
the function $H$ can be expressed in terms of the UV cutoff {$\rho_0$} as
\begin{equation}
\label{Hsol}
{H} =  1+ {g_s\over \pi}
{\frac{m^I d_I}{|\Gamma|}}
\frac{1}{r^4}+ {g_s^2\over 4\pi^2} \frac{
{m^I \widehat C_{IJ}m^J}}{r^4}
\left[\ln\frac{r^4}{{\rho_0^2} {\sigma^2}}-1+
\frac{{\rho^2}}{{\sigma^2}}
\right]~,
\end{equation}
where ${\rho^2=(x^4)^2+(x^5)^2}$,
{$\sigma^2 =\sum_{i=6}^9 (x^i)^2$} and $r^2={\rho^2}+{\sigma^2}$.
At lower scales, the expression is modified as the various types of
twisted fields contribute only above the corresponding enhan\c con.
Without these modifications, the metric exhibits singularities where
{$H=0$}.
\par The flux $\Phi_5(\rho)$ that measures the RR D3-charge contained 
within a surface $\Sigma$ 
comprising  a disk of
radius $\rho$ in   fixed $z$-plane is found to be, 
for $\rho $ above the highest enahn\c con, 
\begin{equation}
\label{flux}{\Phi_5} ({\rho}) =
4\pi^2\,g_s\left({\frac{m^I d_I}{|\Gamma|}}
+{g_s\over 2\pi} {\widehat C_{IJ}m^I m^J}
\ln {\frac{\rho}{\rho_0}}\right)
= 4\pi^2\,g_s\, {Q}({\rho})~,
\end{equation} 
where ${Q}({\rho})={m^I b_I}({\rho})$ is the D3 charge encoded 
in the boundary action. The logarithmic running of the five-form flux is quite
general for fractional brane solutions, and is expected to be related to a
measure of the degrees of freedom of the system. In our case, notice that for
any configuration, we always have $\widehat C_{IJ}m^I m^J\geq 0$, so that the
flux, i.e. the untwisted charge $Q(\rho)$ is always decreasing towards the IR. 
It satisfies the differential equation
\begin{equation}
\label{dQ}
{d {Q}\over d\ln{\rho}} =
{g_s\over 2\pi}\,{\widehat C_{IJ} m^I m^J}
\propto \,{\beta_i  \bar\beta_j\,G^{ij}}~,
\end{equation}
where we introduced the logarithmic derivatives of the twisted scalars,   
${\beta_i} \equiv{d{\tau_i}\over d\ln
{\rho}}$ $= \ii{\widehat C_{iJ} m^J}/(2\pi)$, that is the 
$\beta$-functions for the gauge couplings $\tau_i$, and the metric
$G^{ij}\propto (C^{-1})^{ij}$ appearing in the kinetic terms for the 
twisted scalars themselves, as it can be seen from \eq{sei} and \eq{quattro}. 
The behaviour of the five-form flux has thus a very suggestive formal analogy
with that of the holographic $c$-function of \cite{AGPZ}, which satisfies
$dc/d\ln\rho = 2 \beta_i\bar\beta_j G^{ij}$. Though the latter is defined in the
different context of 5-dimensional gauged supergravity, still we think this
analogy, to be better investigated, supports the possible relation of the flux
to a sensible measure of the degrees of freedom of the system.

\end{document}